\documentclass[11pt]{article}
\newlength{\vshift}
\newlength{\hshift}
\usepackage{amssymb,amsopn}

\newcommand{\newsection}{ \setcounter{equation}{0} \section}

\def\La{\Lambda}
\def\Si{\Sigma}

\def\de{\delta}
\def\be{\beta}
\def\al{\alpha}

\def\ds{\stackrel{\star}{,}}

\providecommand{\href}[2]{#2}

\begin{document}

\begin{titlepage}
\rightline{LMU-TPW 2001-03}
\rightline{MPI-PhT/2001-08}
\rightline{HU-EP-01/13}
\rightline{hep-th/0104153}

\vspace{4em}
\begin{center}

{\Large{\bf Construction of non-Abelian gauge theories on
    noncommutative spaces}}

\vskip 3em

{{\bf B.\ Jur\v co${}^{1}$, L. M\"oller${}^{1,2,3}$, S.\ Schraml${}^{1,3}$, P.\ Schupp${}^{1}$, J.\ Wess${}^{1,2,3}$ }}

\vskip 1em

${}^{1}$Sektion Physik, Universit\"at M\"unchen\\
        Theresienstr.\ 37, D-80333 M\"unchen\\[1em]

${}^{2}$Max-Planck-Institut f\"ur Physik\\
        F\"ohringer Ring 6, D-80805 M\"unchen\\[1em]

${}^{3}$Humboldt-Universit\"at zu Berlin\\
        Institut f\"ur Physik, Invalidenstr. 110, D-10115 Berlin\\

\end{center}

\vspace{2em}

\begin{abstract} 
We present a formalism to explicitly construct non-Abelian gauge
theories on noncommutative spaces (induced via a star product with a
constant Poisson tensor) from a consistency relation. This results
in an expansion of the gauge
parameter, the noncommutative
gauge potential and fields in the fundamental representation, in
powers of a parameter of the noncommutativity. This allows the explicit
construction of actions for these gauge theories. 
\end{abstract}

\vfill

\end{titlepage}\vskip.2cm

\newpage
\setcounter{page}{1}

\newsection{Introduction}

Gauge theories can be formulated on noncommutative spaces. One recent
approach is based on the Seiberg-Witten map \cite{SW}. This is the one
we are particularly interested in because it allows the formulation of
a Lagrangian theory in terms of ordinary fields. One can express such
a theory very intuitively via covariant coordinates  \cite{MSSW}. In this paper we give
an explicit construction for the case of non-Abelian gauge groups. In
contrast to earlier approaches \cite{Bon}, this method now works for
arbitrary gauge groups. 

The idea is to formulate field theories on
noncommutative spaces as theories on commutative spaces and to express the noncommutativity by an
appropriate $\star$-product. Gauge transformations then involve the
$\star$-product as well. This prevents the gauge transformations from being
Lie algebra-valued. They can, however, be defined in the enveloping algebra \cite{JSSW}. It
is possible to find transformations representing the gauge group in the
enveloping algebra that depend on the parameters  and the
gauge potential of the usual gauge theory only. An explicit form of such
transformations can be constructed by a power series expansion in a parameter that
characterizes the noncommutativity. 

In the same manner fields that have the desired $\star$-product
transformation properties can be constructed in terms of fields with the
transformation properties of a usual gauge theory. For the gauge potential
this amounts to the analogue of the Seiberg-Witten map for arbitrary non-Abelian gauge
theories.

Finally we can consider actions that are invariant under the $\star$-product
transformation laws. The Lagrangian of such an action can then be expanded in
terms of the fields of a usual gauge theory and the parameter of the
noncommutativity enters as a coupling constant. New coupling terms for a gauge
theory appear. Such Lagrangians can be seen as effective Lagrangians that are
meaningful in the tree approximation for the description of a
phenomenological S-matrix. But one can also take them serious as
Lagrangians for a quantum field theory with all the radiative corrections. In this context, the renormalizability of the theory has to be investigated \cite{BGPSW1}.

In this paper we explicitly compute the formulas up to second order in the parameter
that characterizes the noncommutativity in the case of the Moyal-Weyl product. 

\newsection{Gauge transformations}

A non-Abelian gauge theory is based on a Lie algebra
\begin{equation}
\label{1}
[T^a,T^b]=if^{ab}_{\hspace{3mm} c} T^c.
\end{equation}

In the usual formulation of a gauge theory fields are considered that transform under gauge transformations 
with Lie algebra-valued infinitesimal parameters\footnote{Throughout we will
  denote fields with this transformation property by $\psi^0$.}: 
\begin{equation}
\label{2}
\de_\al \psi^0(x) =i \al (x)\psi^0(x), \qquad \al(x) = \al_a(x)T^a.
\end{equation}
It follows from (\ref{1}) that
\begin{equation}
\label{3}
(\de_\al \de_\be - \de_\be \de_\al )\psi^0(x)  = i\al_a(x) \be_b(x) f^{ab}_{\hspace{3mm} c} T^c\psi^0(x)
\equiv \de_{\al \times \be}\psi^0(x) ,
\end{equation}
with the shorthand
\begin{equation}
\label{3a}
\al \times \be \equiv \al_a \be_b f^{ab}_{\hspace{3mm} c} T^c =-i[\al,\be].\nonumber
\end{equation}

In addition, a gauge potential $a_{i,a}(x)$ is introduced with the transformation property
\begin{equation}
\label{4}
\de_\al a_{i,a}(x) = \partial_i\al_a(x)-f^{bc}_{\hspace{3mm} a}\al_b(x)a_{i,c}(x),
\end{equation}
or equivalently,
\begin{eqnarray}
\label{5}
a_i(x) & =& a_{i,a}(x)T^a \\
\de_\al a_i(x) & = & \partial_i \al(x) +i[\al(x),a_i(x)]. \nonumber
\end{eqnarray}
This allows the definition of covariant derivatives and the field strength:
\begin{eqnarray}
\label{6}
{\cal{D}}_i \psi^0(x) & =& (\partial_i-ia_i)\psi^0(x) \\
F^0_{ij}\psi^0(x)  & = & i\big({\cal{D}}_i{\cal{D}}_j-{\cal{D}}_j{\cal{D}}_i\big) \psi^0(x). \nonumber
\end{eqnarray}

In a gauge theory on noncommutative coordinates (\ref{2}) is replaced by
\begin{equation}
\label{7}
\de_\La \psi(x) =i \La (x)\star\psi(x).
\end{equation}
The $\star$-product based on a quite general coordinate algebra has
been defined in \cite{JSSW}. In this paper, we shall evaluate the
respective formulas for the Moyal-Weyl-product only. This is the
product that is most frequently discussed in the current literature, but we
emphasize that the methods used in this paper work for other
$\star$-products as well.

The $\star$-product of two functions does not commute, it reflects the
algebraic properties of the space coordinates. As a consequence, two
transformations of the type (\ref{7}) cannot be reduced to the matrix commutator 
of the generators of the Lie algebra:
\begin{equation}
\label{8}
(\de_\La\de_\Si-\de_\Si\de_\La) \psi(x) = \big(\La
(x)\star\Si(x)-\Si(x)\star\La(x)\big)\star\psi(x) \equiv [\La (x)\ds\Si(x)]\star\psi(x).
\end{equation}
The parameters cannot be Lie algebra-valued, they  have to be in the enveloping algebra:
\begin{equation}
\label{9}
\La(x) = \La_a(x)T^a + \La^1_{ab}(x) : T^a T^b: + \dots + \La^{n-1}_{a_1
  \dots a_{n}}(x) : T^{a_{1}}\dots T^{a_n} : + \dots \hspace{3mm}  .
\end{equation}

The dots indicate that we have to sum over a basis of the vector space spanned by 
the homogeneous polynomials in the generators of the Lie algebra. Completely symmetrized products 
form such a basis:
\begin{eqnarray}
\label{10}
:T^a: & =& T^a \\
:T^aT^b: & = & \frac{1}{2}\{T^a,T^b\}= \frac{1}{2}(T^aT^b+T^bT^a)\nonumber \\
:T^{a_1}\dots T^{a_n}: & = & \frac{1}{n!}\sum_{\pi \in S_n}T^{a_{\pi(1)}}\dots T^{a_{\pi(n)}}.\nonumber
\end{eqnarray}

The $\star$-commutator of two enveloping algebra-valued transformations will 
remain enveloping algebra-valued. The price we seem to have to pay are
infinitely many parameters $\La_{a_1\dots a_n}^{n-1}(x)$, however, it is possible to define gauge transformations where all these infinitely 
many parameters depend  on the usual gauge parameter $\al(x)$ and the
gauge potential $a_i(x)$ and on their
derivatives. Transformations of this type will be denoted
$\La_{\al}[a] $ and their $x$-dependence is purely via this
finite set of parameters and gauge potentials $\La_{\al}[a] \equiv
\La_{\al(x)}[a(x)] $ (for constant $\theta$).

The transformation (\ref{7}) will now be restricted to such  parameters $\La_{\al}[a]$
\begin{equation}
\de_{\al} \psi(x) =i \La_{\al}[a]\star\psi(x).
\label{13}
\end{equation}
Each finite set of parameters $\al_a(x)$ defines a ``tower'' $\La_{\al}[a]$ in 
the enveloping algebra that is entirely determined by the Lie
algebra-valued part. To define and construct this tower we demand in
accord with (\ref{3}) (cf. \cite{JSSW}): 
\begin{equation}
\label{14}
(\de_{\al} \de_{\be} - \de_{\be} \de_{\al} )\psi(x)  = \de_{\al \times \be}\psi(x). 
\end{equation}
The variations $\de_{\al}$ are those of equation (\ref{13}).
More explicitly:
\begin{equation}
\label{15}
i\de_{\al}\La_{\be}[a]-i\de_{\be}\La_{\al}[a]+\La_{\al}[a]\star\La_{\be}[a]-\La_{\be}[a]\star\La_{\al}[a]
= i\La_{\al\times \be}[a]. 
\end{equation}
The variation  $\de_{\be}\La_{\al}[a]$ refers to the 
$a_i$-dependence of $\La_{\al}[a]$ and the transformation property (\ref{4}) of $a_i$. 

 It is natural to expand the star product 
in its ``noncommutativity'' and to solve (\ref{15}) in a power series expansion.
For this purpose we introduce a parameter $h$:
\begin{eqnarray}
\label{16}
\big(f \star g\big)(x)& =&e^{\frac{i}{2}h\frac{\partial}{\partial
    x^i}\theta^{ij}\frac{\partial}{\partial y^j}}f(x)g(y)|_{y\rightarrow x}\\
& = &f(x)g(x)+\frac{i}{2}h\theta^{ij}\partial_i f(x)\partial_j g(x)-\frac{1}{8}h^2\theta^{ij}\theta^{kl}\partial_i\partial_k f(x)\partial_j\partial_l g(x)+\dots\hspace{3mm}.\nonumber
\end{eqnarray}
We could have used $\theta$ as an expansion parameter, however a $\theta$-dependence 
of the fields might and will in fact arise for other reasons.
 
We assume that it is possible to expand the tower $\La_{\al}[a]$ in
the parameter $h$:
\begin{equation}
\label{16a}
\La_{\al}[a] = \al + h \La^1_{\al}[a] + h^2 \La^2_{\al}[a]+
\cdots \hspace{3mm} .
\end{equation}
Now we expand equation (\ref{15}) in $h$. To zeroth order  we find equation (\ref{3}) 
which has the solution (\ref{2}). To first order we obtain
\begin{equation}
\label{17}
i\de_{\al}\La^1_{\be}[a]-i\de_{\be}\La^1_{\al}[a]+[\al,\La^1_{\be}[a]]-[\be,\La^1_{\al}[a]]-i\La^1_{\al\times\be}[a]=
-\frac{i}{2}\theta^{ij}\{\partial_i\al,\partial_j\be\} .
\end{equation}
There is a homogeneous part in $\La^1_{\al}[a]$ and an inhomogeneous
part. We solve the inhomogeneous part
with an ansatz linear in $\theta$, because 
the inhomogeneous part is linear in $\theta$ as well. For dimensional reasons 
there is only a finite number of terms that can be used in such an ansatz. 
The proper combination of such terms is
\begin{equation}
\label{18}
\La^1_{\al}[a]=\frac{1}{4}\theta^{ij}\{\partial_i\al,a_j\}
=\frac{1}{2}\theta^{ij}\partial_i\al_a a_{j,b} :T^aT^b: .
\end{equation}
The derivative term $\partial_i\al$ in the variation of the gauge potential (\ref{5}) compensates the 
inhomogeneous part in (\ref{17}), whereas the commutator term of the variation 
of the gauge potential combines with other terms of (\ref{17}) to produce 
$\La^1_{\al\times\be}[a]$. 

We can add solutions of the homogeneous part of (\ref{17}). 
If there is a  quantity $F_i$ with the covariant transformation property
\begin{equation}
\label{19}
\de_{\al}F_i=i[\al,F_i]
\end{equation}
and such a term can easily be constructed, for instance from the field strength and the covariant derivative in (\ref{6})
\begin{equation}
\label{22}
F_i=\theta^{jk}{\cal{D}}_jF_{ki}^0,
\end{equation}
then we find a solution $\widetilde{\La}^1_{\al}[a]$ of the homogeneous part of 
equation (\ref{17}):
\begin{equation}
\label{20}
\widetilde{\La}^1_{\al}[a]=c\theta^{ij}\{\partial_i\al,F_j\}.
\end{equation}
This term can be added to $\La^1_{\al}[a]$ defined in (\ref{18}). To first order in 
$h$ we obtain
\begin{equation}
\label{21}
\La^1_{\al}[a]=\theta^{ij}\{\partial_i\al,\frac{1}{4}a_j+cF_j\}.
\end{equation}

We can  view the additional term as a redefinition of the gauge potential:
\begin{equation}
\label{23}
\widetilde{a_i}=a_i+4cF^{\al}_i.
\end{equation}
It does not change the transformation properties (\ref{5}) 
\begin{equation}
\label{24}
\de_{\al}\widetilde{a_i}=\partial_i\al +i[\al,\widetilde{a_i}].
\end{equation}
Such generalized solutions as (\ref{21}) have to be expected, as it is only 
the transformation property of $a_i$ that matters. To define a
physical theory, we have to make use of the full freedom of the
Seiberg-Witten map. It has been shown that for finding a renormalizable
theory the extra terms are essential \cite{BGPSW3}.

There are other solutions of the homogeneous equation that cannot be
obtained from a redefinition of the vector potential $a_i$. An example is
\begin{equation}
\label{24a}
\widetilde{\La}^{1\hspace{1mm}'}_{\al}[a]=c\theta^{ij}[\partial_i\al,a_j].
\end{equation}

The choice of the constant $c=-\frac{1}{4}$ for this solution of the homogeneous part
of (\ref{17}) would  simplify some of the calculations to
come, however, we decide to work with  (\ref{18}) instead,
since this  is a solution expressed in the completely symmetrized basis (\ref{10}) in the
generators of the Lie algebra.

To second order in $h$ we obtain from (\ref{15}):
\begin{eqnarray}
\label{25}
&&i\de_{\al}\La^2_{\be}[a]-i\de_{\be}\La^2_{\al}[a]+[\al,\La^2_{\be}[a]]-[\be,\La^2_{\al}[a]]-i\La^2_{\al\times\be}[a]=\\
&&\hspace{3mm}+\frac{1}{8}\theta^{ij}\theta^{kl}[\partial_i\partial_k\al,\partial_j\partial_l\be]-\frac{i}{2}\theta^{ij}\Big(\{\partial_i\La^1_{\al}[a],\partial_j\be\}
-\{\partial_i\La^1_{\be}[a],\partial_j\al\}\Big)-
[\La_\al^1[a],\La_\be^1[a]].\nonumber
\end{eqnarray}
The homogeneous part of the equation has the same structure as before.  We
shall use the expression (\ref{18}) for $\La^1_\al[a]$ and we see that
the terms of the inhomogeneous part involving $\La^1_\al[a]$
contribute to third order in $T^a$.  With an appropriate ansatz we can eliminate all
these terms of third order and of second order in $T^a$ as well. The respective terms in the solution (\ref{26}) can easily be identified. Finally we obtain a solution of (\ref{25})\footnote{Similar
  results have been obtained in \cite{GH} and \cite{AK} in the context
  of $U(n)$.}:
\begin{eqnarray}
\label{26}
\La_\al^{2}[a]&=&\frac{1}{32}\theta^{ij}\theta^{kl}\Big(-4\{\partial_{i}\alpha,\{a_{k},\partial_{l}a_{j}\}\}-i\{\partial_{i}\alpha,\{a_{k},[a_{j},a_{l}]\}\}-i\{a_{j},\{a_{l},[\partial_{i}\alpha,a_{k}]\}\}\nonumber\\
&&+2i[\partial_{i}\partial_{k}\alpha,\partial_{j}a_{l}]-2[\partial_{j}a_{l},[\partial_{i}\alpha,a_{k}]]+2i[[a_{j},a_{l}],[\partial_{i}\alpha,a_{k}]]\Big).
\end{eqnarray}

The solutions (\ref{18}) and (\ref{26}) are such that they are of first and second
order in $\theta$ respectively. We know from \cite{JSSW} that we can expect a
solution of (\ref{15}) where the order in $\theta$ and the order in $T^a$ are
related. In such a solution the contribution in $\theta^n$ will be of order
$n+1$ in $T^a$. The above solutions are of this type. This can however be changed by adding
$\theta$-dependent solutions of the homogeneous equation.

\newsection{Fields}
In a usual gauge theory, fields have the transformation property
(\ref{2}). We have denoted fields that transform this way by
$\psi^0$. In a gauge 
theory with the $\star$-product fields are supposed to transform as in
(\ref{13}). We show that fields with this transformation property can be built
from fields with the transformation property (\ref{2}) and the gauge potential
$a_{i}$. 

We expand  in powers of $h$:
\begin{equation}
\label{27}
\psi[a]=\psi^0 +h\psi^1[a]+h^2\psi^2[a]+ \cdots \hspace{3mm}.
\end{equation}
To zeroth order in $h$, we obtain (\ref{2}) and to first order:
\begin{equation}
\label{28}
\de_\al\psi^1[a]=i\al\psi^1[a]
+i\La_\al^1[a]\psi^0-\frac{1}{2}\theta^{ij}\partial_i\al\partial_j\psi^0.
\end{equation}
If we take the solution (\ref{18}) for $\La_{\al}^1[a]$, we find that
\begin{equation}
\label{29}
\psi^1[a]=-\frac{1}{2}\theta^{ij}a_i\partial_j\psi^0+\frac{i}{4}\theta^{ij}a_ia_j\psi^0.
\end{equation}
will have the desired transformation property (\ref{13}) to first order in $h$.
We proceed to the next order,
\begin{eqnarray}
\label{29a}
&&\de_\al\psi^2[a]=i\al\psi^2[a]+i\La_\al^1[a]\psi^1[a]
+i\La_\al^2[a]\psi^0-\frac{1}{2}\theta^{ij}\partial_i\La^1_\al[a]\partial_j\psi^0\\
&&\hspace{12mm}-\frac{1}{2}\theta^{ij}\partial_i\al\partial_j\psi^1[a]-\frac{i}{8}\theta^{ij}\theta^{kl}\partial_i\partial_k\al\partial_j\partial_l\psi^0,\nonumber
\end{eqnarray}
use (\ref{26}) for $\La_{\al}^2[a]$ and find:
\begin{eqnarray}
\label{30}
\psi^{2}[a]&=&\frac{1}{32}\theta^{ij}\theta^{kl}\Big(
-4i\partial_{i}a_{k}\partial_{j}\partial_{l}\psi^0
+4a_{i}a_{k}\partial_{j}\partial_{l}\psi^0
+8a_{i}\partial_{j}a_{k}\partial_{l}\psi^0\\
&& 
-4 a_{i}\partial_{k}a_{j}\partial_{l}\psi^0
 -4i a_{i}a_{j}a_{k}\partial_{l}\psi^0
+4i a_{k}a_{j}a_{i}\partial_{l}\psi^0-4ia_{j}a_{k}a_{i}\partial_{l}\psi^0\nonumber\\
&&+4 \partial_{j}a_{k}a_{i}\partial_{l}\psi^0
-2\partial_{i}a_{k}\partial_{j}a_{l}\psi^0
+4ia_{i}a_{l}\partial_{k}a_{j}\psi^0
+4ia_{i}\partial_{k}a_{j}a_{l}\psi^0
\nonumber\\
&&-4ia_{i}\partial_{j}a_{k}a_{l}\psi^0
+3a_{i}a_{j}a_{l}a_{k}\psi^0
+4a_{i}a_{k}a_{j}a_{l}\psi^0
+2a_{i}a_{l}a_{k}a_{j}\psi^0\Big).\nonumber
\end{eqnarray}
Homogeneous solutions of (\ref{28}) and (\ref{28}) can naturally be
added to these solutions.
The adjoint field $\bar{\psi}[a]$ is easily obtained from (\ref{29})
and (\ref{30}), keeping in mind that $a_{i}$ is supposed to be
self-adjoint for a unitary gauge group.

\newsection{Gauge potentials and field strengths}
In the same way as in the last section for ordinary fields we can solve for an enveloping
algebra-valued gauge potential. Its transformation property is (for a
definition, e.g. \cite{MSSW}):
\begin{equation}
\label{31}
\de_\al A_i=\partial_i\La_\al[a] +i[\La_\al[a] \ds A_i].
\end{equation}
Again we expand in $h$:
\begin{equation}
\label{32}
A_i[a]=a_i +hA_i^1[a]+h^2A^2_i[a]+ \cdots \hspace{3mm}.
\end{equation}
As expected, we can recapture (\ref{5}) to zeroth order, to first order we obtain:
\begin{equation}
\label{33}
\de_\al
A^1_i[a]=\partial_i\La_\al^1[a]+i[\La_\al^1[a],a_i]+i[\al,A^1_i[a]]-\frac{1}{2}\theta^{kl}\{\partial_k\al,\partial_l
a_i\}.
\end{equation}
Again we use the solution (\ref{18}) for $\La_{\al}^1[a]$ and find as a solution
to (\ref{33})
\begin{equation}
\label{34}
A^1_i[a]=-\frac{1}{4}\theta^{kl}\{a_k,\partial_l a_i+F^0_{li}\},
\end{equation}
where $F^0_{ij}$ is the field strength of the ordinary Lie algebra-valued gauge
theory, introduced in (\ref{6})
\begin{equation}
\label{35}
F^0_{ij}=\partial_i a_j-\partial_j a_i - i [a_i,a_j].
\end{equation}
To second order in $h$ we obtain from (\ref{31}):
\begin{eqnarray}
\label{35a}
 \delta_{\alpha}A_{i}^{2}[a]&=& \partial_i\La_{\al}^{2}[a] +i [\al,A_i^{2}[a]]
 + i[\Lambda_\al^{1}[a],A_i^{1}[a]] + i[\La_\al^{2}[a],a_i]-\\
 &&-\frac{1}{2}\theta^{kl}\{\partial_{k}\alpha,\partial_{l}A_i^{1}[a]\}-\frac{1}{2}\theta^{kl}\{\partial_{k}\Lambda_\al^{1}[a],\partial_{l}a_i\}-\frac{i}{8}\theta^{kl}\theta^{mn}[\partial_{k}\partial_{m}\alpha,\partial_{l}\partial_{n}a_i].\nonumber
\end{eqnarray}
With the choice (\ref{26}) for $\La_{\al}^2[a]$ this has the following solution:
\begin{eqnarray}
\label{36}
A_{i}^{2}[a]&=&\frac{1}{32}\theta^{kl}\theta^{mn}\Big(
4i[\partial_k\partial_ma_i,\partial_l a_n]
-2i[\partial_k\partial_ia_m,\partial_l a_n]
+4\{a_k,\{a_m,\partial_n F^0_{li}\}\}\\
&& 
+2[[\partial_k a_m,a_i],\partial_l a_n]
-4\{\partial_l a_i,\{\partial_m a_k,a_n\}\}
+4\{a_k,\{F^0_{lm},F^0_{ni}\}\}\nonumber\\
&&-i\{\partial_i a_n,\{a_l,[a_m,a_k]\}\}
-i\{a_m,\{a_k,[\partial_i a_n,a_l]\}\}\nonumber\\
&&
+4i[[a_m,a_l],[a_k,\partial_n a_i]]
-2i[[a_m,a_l],[a_k,\partial_i a_n]]
-\{a_m,\{a_k,[a_l,[a_n,a_i]]\}\}\nonumber\\
&&+\{a_k,\{[a_l,a_m],[a_n,a_i]\}\}
+[[a_m,a_l],[a_k,[a_n,a_i]]]\Big).
\nonumber
\end{eqnarray}

With this solution  at hand we now  turn to  the enveloping
algebra-valued field strength (defined in \cite{MSSW}):
\begin{equation}
\label{37}
F_{ij}=\partial_i A_j-\partial_j A_i - i [A_i \ds A_j],
\end{equation}
with the transformation property
\begin{equation}
\label{37a}
\de_\al F_{ij}=i[\La_\al[a] \ds F_{ij}].
\end{equation}
To express this field strength, we insert (\ref{34}) and (\ref{36})
into (\ref{37}). We could have used (\ref{37a}) to find a field with the
desired transformation property, as we did in section 3. This however
does not reproduce the full solution (\ref{38}) which rests on the
definition of $F_{ij}[a]$ in terms of the gauge potential (\ref{32}):
\begin{equation}
\label{38}
F^1_{ij}[a]=\frac{1}{2}\theta^{kl}\{F^0_{ik},F^0_{jl}\}-\frac{1}{4}\theta^{kl}\{a_k,(\partial_l+{\cal{D}}_l)F^0_{ij}\}.
\end{equation}
$F^2_{ij}[a]$ to second order in $h$ is obtained similarly by
inserting (\ref{36}) into (\ref{37}).
The gauge potential $A_i$ with its transformation property (\ref{31}) allows the definition of a covariant derivative  
\begin{equation}
\label{38a}
{\cal{D}}_i\psi = \partial_i\psi -i A_i \star \psi
\end{equation}
with the transformation (\ref{13}):
\begin{equation}
\label{38b}
\de_\al{\cal{D}}_i\psi = i\La_\al[a]\star{\cal{D}}_i\psi.
\end{equation}

\newsection{Actions}
The transformation laws of the field strength (\ref{37a}), the fields (\ref{13}) and
the covariant derivatives (\ref{38b}) allow  the construction of invariant
actions. It can be shown by partial integration that the integral has the
trace property for the $\star$-product:
 \begin{equation}
\label{39}
\int f \star g \hspace{2mm}\textrm{d}x = \int g \star f \hspace{2mm}\textrm{d}x = \int f g \hspace{2mm}\textrm{d}x.
\end{equation}

Thus we find an invariant  action for the gauge potential
\begin{equation}
\label{40}
S= -\frac{1}{4}\textrm{Tr}\int F_{ij} \star F^{ij} \hspace{2mm}\textrm{d}x,  
\end{equation}
as well as for the matter fields
\begin{equation}
\label{40a}
S=\int  \bar{\psi}\star (\gamma^i {\cal{D}}_i-m) \psi  \hspace{2mm}\textrm{d}x.
\end{equation}

Our aim is to expand these actions in the fields $a_i$ and $\psi^0$ and to
treat them as conventional field theories depending on a coupling constant
$\theta$. We only do this here to first order in $h$ and construct 
the Lagrangian from our previous results:
\begin{eqnarray}
\label{41}
m\bar{\psi}\star\psi&=& m\bar{\psi}^0\psi^0
+\frac{i}{2}h\theta^{kl}m{\cal{D}}_k\bar{\psi}^0{\cal{D}}_l\psi^0\nonumber \\
\bar{\psi}\star\gamma^i{\cal{D}}_i\psi&=&\bar{\psi}^0\gamma^i{\cal{D}}_i\psi^0
+\frac{i}{2}h\theta^{kl}{\cal{D}}_k\bar{\psi}^0\gamma^i{\cal{D}}_l{\cal{D}}_i\psi^0
-\frac{1}{2}h\theta^{kl}\bar{\psi}^0\gamma^iF^0_{ik}{\cal{D}}_l\psi^0\nonumber \\   
F_{ij}\star F^{ij}&=&F_{ij}^0
F^{0ij}+\frac{i}{2}h\theta^{kl}{\cal{D}}_k F_{ij}^0{\cal{D}}_l F^{0ij}+\frac{1}{2}h\theta^{kl}\{\{F_{ik}^0,F_{jl}^0\},F^{0ij}\}  \nonumber\\
&&-\frac{1}{4}h\theta^{kl}\{F_{kl}^0,F_{ij}^0F^{0ij}\}-\frac{i}{4}h\theta^{kl}[a_k,\{a_l,F_{ij}^0F^{0ij}\}]\nonumber
\end{eqnarray}

For the action we use partial integration and the cyclicity of the
trace and obtain to first order in $h$:
\begin{eqnarray}
\label{42}
\int \bar{\psi}\star(\gamma^i {\cal{D}}_i-m)\psi\hspace{1mm}\textrm{d}x&=& \int
 \bar{\psi}^0(\gamma^i {\cal{D}}_i-m)\psi^0\hspace{1mm}\textrm{d}x -\frac{1}{4}h\theta^{kl}\int
 \bar{\psi}^0 F_{kl}^0(\gamma^i {\cal{D}}_i-m)\psi^0\hspace{1mm}\textrm{d}x\nonumber \\
&&-\frac{1}{2}h\theta^{kl}\int
 \bar{\psi}^0\gamma^i F_{ik}^0 {\cal{D}}_l\psi^0\hspace{1mm}\textrm{d}x \\
-\frac{1}{4}\textrm{Tr}\int F_{ij} \star
 F^{ij}\hspace{1mm}\textrm{d}x&=&-\frac{1}{4}\textrm{Tr}\int F^0_{ij} F^{0 ij}\hspace{1mm}\textrm{d}x+\frac{1}{8}h\theta^{kl}\textrm{Tr}\int F^0_{kl}F^0_{ij} F^{0 ij}\hspace{1mm}\textrm{d}x \nonumber\\
&&-\frac{1}{2}h\theta^{kl}\textrm{Tr}\int F^0_{ik}F^0_{jl} F^{0 ij}\hspace{1mm}\textrm{d}x
\end{eqnarray}

\section{The Abelian case}
Noncommutative Abelian gauge theories have recently been studied intensively and substantial results have been obtained.

If such a theory is expanded not in the
noncommutativity $h$ as in the previous chapters, but in powers of the
gauge potential of the commutative theory as suggested in \cite{MW}, the following result is
valid to all orders in $\theta$ and first non-trivial order in $a$:
\begin{eqnarray}
\label{43}
A^i \lbrack a \rbrack &=& \theta^{ij}\big(a_j + \frac{1}{2}\theta^{kl}a_l \star_2
(\partial_k a_j +f_{kj}) + \cdots \big) \\
\Lambda_\al \lbrack a \rbrack &=& \al +\frac{1}{2}\theta^{kl}a_l \star_2
\partial_k \al + \cdots\hspace{3mm},
\end{eqnarray}
where $f_{jk}=\partial_j a_k - \partial_k a_j$ is the Abelian field
strength and $\star_2$ is an abbreviation for the following power series in
the noncommutativity\footnote{This notation $\star_2$ is now widely
  used, e.g. in \cite{MW}, \cite{Liu} and \cite{Oku}.} (it is not a $\star$-product though):
\begin{equation}
\label{44}
f \star_2 g = \mu \Big( \frac{\sin(\frac{h \theta^{ij}}{2}\partial_i
  \otimes \partial_j )}{\frac{h \theta^{ij}}{2}\partial_i
  \otimes \partial_j }\Big)(f\otimes g),
\end{equation}
and $\mu(f\otimes g)=f \cdot g$ the ordinary
multiplication map.  It is particularly convenient to use this
multiplication in the Fourier representation.

We will now derive this result.
We know from \cite{JSSW} and \cite{JSW} that the following expressions
for the noncommutative gauge potential and gauge parameter satisfy
both the Seiberg-Witten gauge condition and the consistency relation
(these expressions are valid for arbitrary Poisson
structures $\theta(x)$):
\begin{eqnarray}
\label{45}
A^i\lbrack a \rbrack &=& \big(\exp(a_\star +\partial_t)-1\big)\hspace{1mm} x^i \\
\Lambda_\al \lbrack a \rbrack &=& \Big(\frac{\exp(a_\star +\partial_t)-1}{a_\star +\partial_t}\Big)\hspace{1mm}\al,
\end{eqnarray}
with the differential operator
\begin{equation}
\label{46}
a_\star = \sum
\frac{(ih)^n}{n!}U_{n+1}(a_\theta,\theta,\cdots ,\theta)=\theta^{ij}a_j
\partial_i + \cdots\hspace{3mm},
\end{equation}
and the rule $\partial_t
\theta^{il}=-\theta^{ij}f_{jk}\theta^{kl}$. The differential operator
$a_\star$ is obtained from the vector field $a_\theta = \theta^{ij}
a_j \partial_i$ and the Poisson bivector $\theta=\theta^{ij}\partial_i
\otimes \partial_j$ via Kontsevich's formality maps $U_n$ (for further
clarifications we refer to the mentioned articles). 

Expanding the
exponentials results in an expansion in powers of the ordinary gauge
potential $a_i$, each term containing \emph{all} powers of $h$:
\begin{eqnarray}
\label{47}
A^i\lbrack a \rbrack &=& \underbrace{a_\star x^i}_{\mathcal{O}(a^1)}
+\underbrace{\frac{1}{2}(a_\star^2 + \dot{a}_\star)
  x^i}_{\mathcal{O}(a^2)} +\cdots \\
\Lambda_\al \lbrack a \rbrack &=&
\underbrace{\al}_{\mathcal{O}(a^0)}+\underbrace{\frac{1}{2}a_\star \al}_{\mathcal{O}(a^1)} + \cdots\hspace{3mm}.
\end{eqnarray}
Kontsevich has given a graphical prescription similar to Feynman
diagrams to compute the formality maps. Using these it is
straightforward to compute
$a_\star$ explicitly to all orders in $h$ for constant $\theta$. The result is:
\begin{equation}
\label{48}
a_\star g = \big(\theta^{ij} a_j\big) \star_2 \partial_i g= \theta^{ij} a_j
\partial_i g + \cdots \hspace{3mm},
\end{equation}
with the already mentioned product $\star_2$. Inserting this $a_\star$
into (\ref{47}) and using the fact that $a_\star x^i = a_\theta x^i$ gives the expressions for $A^i [a]$ and $\La_\al [a]$
stated at the beginning of this section. Higher order terms can be
obtained similarly. A nice heuristic derivation of these results based
on the consistency condition has been given in \cite{Oku}.

\section{Expansion of non-Abelian fields in $a$}
Adopting a similar approach like in the  previous section, we here
state a straight-forward result concerning
the expansion of fields in a non-Abelian gauge theory in powers of the
commutative gauge potential. 

Assume that a field $\psi$ (a representation of the enveloping algebra
$\de_\al \psi = i \La_\al[a] \star \psi$) can
be written as a matrix-valued differential operator $\Phi[a]$ applied to the
field $\psi^0$ in the representation of the Lie algebra
($\de_\al\psi^0=i\al \psi^0$): $\psi = \Phi[a]
\psi^0$. 
Then the variation of $\psi$ can be written in the following way:
\begin{equation}
\label{49}
\big(\de_\al \Phi[a]\big)\psi^0 + \Phi[a] \big(i \al \psi^0\big)
\stackrel{!}{=} i \La_\al[a] \star \big(\Phi[a]\psi^0\big).
\end{equation}
To zeroth order in $a$ this reads:
\begin{equation}
\label{50}
\Phi^1[\partial \al]\psi^0 +  i \al \psi^0 = i \al \star \psi^0.
\end{equation}
The second term in the variation of $a$, $i[\al,a]$, drops out being
first order in $a$. Due to the Bianchi identity of a non-Abelian gauge
theory ($df + a\wedge f=0$), this  expansion is not well defined to
higher orders in $a$ and we will not discuss orders different from
$\mathcal{O}(a^0)$.  This problem does not occur for the Abelian case.
Continuing in our analysis we set
\begin{equation}
\label{51}
\Phi^1[\partial \al]\psi^0 =   i \al \star \psi^0 - i \al \psi^0 =:
-\frac{h}{2} \theta^{kl}(\partial_k \al) \bullet \partial_l \psi^0,
\end{equation}
where we have introduced the following shorthand\footnote{This product
  was also introduced in \cite{Gar}, there called $\star''$.}
\begin{equation}
\label{52}
f \bullet g =
\mu\Big(\frac{e^{\frac{ih}{2}\theta^{kl}\partial_k\otimes\partial_l}-1}{\frac{ih}{2}\theta^{kl}\partial_k\otimes\partial_l}\Big)(f\otimes
g).
\end{equation}
This should be compared with the Moyal-Weyl-product: $f\star
g:=\mu(e^{\frac{ih}{2}\theta^{kl}\partial_k\otimes\partial_l})(f\otimes g)$.
With this shorthand for the product we are free to integrate:
\begin{equation}
\label{53}
\Phi^1[a_k]\psi^0 =  
-\frac{h}{2} \theta^{kl}(a_k) \bullet \partial_l \psi^0.
\end{equation}
Therefore we obtain the following expansion (to first power in
$a$ and all powers in $h$):
\begin{equation}
\label{54}
\psi =  \psi^0
-\frac{h}{2} \theta^{kl}(a_k) \bullet \partial_l \psi^0 + \cdots \hspace{3mm},
\end{equation}
Okuyama \cite{Oku} has computed $A^i[a]$ and $\La_\al[a]$
in a similar fashion.

\section*{Acknowledgement}
We thank Dieter L\"ust for his kind hospitality and inspiring
  discussions. Also we thank B. Zumino et al.~\cite{Zu} for drawing
  our attention to a previous mistake in (\ref{26}). A previous misprint in
  (\ref{42}) was pointed out by C.E. Carlson et al.~\cite{Ca}. 


\end{document}